# Topological sorting of magnetic colloidal bipeds†

Aneena Rinu Perayil,[a] Piotr Kuświk,[b] Maciej Urbaniak,[b] Feliks Stobiecki,[b] Sapida Akhundzada,[c] Arno Ehresmann,[c] Daniel de las Heras[d] and Thomas M. Fischer *[a]

Topologically nontrivial adiabatic loops of the orientation of a homogeneous external magnetic field drive the walking of paramagnetic colloidal bipeds above a deformed quasi-periodic magnetic square pattern. Depending on the topological properties of the loop we can simultaneously control the walking directions of colloidal bipeds as a function of their size and as a function of the size of a deformed unit cell of the pattern. The bipeds walk performing steps with their two feet alternatingly grounding one foot and lifting the other. The step width of the bipeds is given by a set of winding numbers $(w_x, w_y) \in \mathbb{Z}^2$ – a set of topological invariants – that can only change by integers as we continuously increase the length of the bipeds. We experimentally use this discrete size dependence for the robust sorting of bipeds according to their length.

## 1. Introduction

In a first step, sorting classifies a set of objects into disjunct categories that are determined by the properties of the objects. In a second step, subsets of objects falling into one category are transported to a region in space reserved for one category thereby being separated in space from objects belonging to other categories. The properties of the objects can be discrete. For example, in the Golgi apparatus different proteins are sorted and packed into vesicles for transport to various destinations within the cell.[1] On the cellular scale, during the embryo genesis the cells differentiate and move toward their correct location.[2,3] The objects properties can also be continuous such as the length,[4] the width,[5] and the size[6–11] among others.[11–14] In the continuous case, categories are intervals and the interval boundaries are bifurcation points between different discrete categories. Sorting objects with a computer algorithm is a serial or weakly parallel process, while the separation of chemicals *via* electrophoresis is a massive parallel process. Massive parallel processes allow for high-throughput sorting.[15] Active sorting[16–19] occurs when self propelled objects sort themselves. Active sorting is usually more robust than passive sorting.[20,21]

Microfluidic sorting devices often rely on optical,[22] steric[23] patterns that guide specific particles along different conduits. In biology one tries to understand sorting mechanisms – originally developed *via* evolution – by analyzing whether the sorting persists or not after applying all kinds of perturbations.[24–27]

Analytical[24–27] as well as machine learning[28,29] approaches try to identify and exploit how a preexisting sorting mechanism works. Synthetic approaches are used to construct a sorting process by having understood *a priori* how it works. Here we follow a synthetic path to generate a sorting device for colloidal bipeds of different length from first principles. Knowing the mathematical laws that govern the transport of colloidal bipeds[30–32] we design a magnetic pattern and a modulation that forces the sorting of colloidal bipeds according to their length using external commands.

The mathematical principles that govern the transport are of topological nature, that is, the transport is fully controlled by a set of topological invariants. We have used those topological laws to robustly transport paramagnetic and diamagnetic colloids on periodic patterns,[33–35] to independently transport them into different directions,[33,34] to simultaneously write different letters,[30] or to walk in a time reversal or non-time reversal way.[36] We have induced skipping orbits at the edge of topologically distinct[37] and topologically equivalent[38] lattices. We have generalized this topological concept to non-periodic metamorphic patterns that are locally but not globally periodic[39] and used the concept to synthesize colloidal assemblies called bipeds of a desired length on a hexagonal pattern with constant lattice constant but varying symmetry phase[32] and to cloak certain regions against colloidal trespassers.[40] In this work we show that a metamorphic square pattern with varying lattice constant and a magnetic driving loop with a mirror symmetry different from ref. 32 can force the colloids to automatically fulfill a new task compared to those reported in ref. 30–40: different length colloidal bipeds sort themselves. This requires them not only to separate but additionally to walk to predefined attractor locations differing for each different biped length.

[a] *Experimentalphysik X, Physikalisches Institut, Universität Bayreuth, D-95440 Bayreuth, Germany. E-mail: thomas.fischer@uni-bayreuth.de*
[b] *Institute of Molecular Physics, Polish Academy of Sciences, 60-179 Poznań, Poland*
[c] *Institute of Physics and Center for Interdisciplinary Nanostructure Science and Technology (CINSaT), Universität Kassel, D-34132 Kassel, Germany*
[d] *Theoretische Physik II, Physikalisches Institut, Universität Bayreuth, D-95440 Bayreuth, Germany*

† Electronic supplementary information (ESI) available. See DOI: https://doi.org/10.1039/d4sm01480d

## 2. Setup

Paramagnetic colloidal particles (diameter $d = 2.8$ μm) immersed in water are placed on top of a two-dimensional magnetic pattern. The pattern is a deformed square lattice of alternating regions with positive and negative magnetization relative to the direction normal to the pattern, see Fig. 1. For our thin film we choose a pattern with magnetization

$$\mathbf{M}(\mathbf{r}_\mathcal{A}, z) = M_s \delta\left(\frac{z}{h}\right) \mathbf{n}\, \mathrm{sign}\left(\sum_{i=0}^{1} \cos(\mathbf{q}_i \cdot \mathbf{r}_\mathcal{A})\right), \quad (1)$$

where $\mathbf{r}_\mathcal{A}$ is the two dimensional vector in action space $\mathcal{A}$, which is the plane parallel to the pattern where the colloids move. The modulus of the saturation magnetization is denoted by $M_s$, $h$ is the thickness of the film, $\mathbf{n}$ is the vector normal to the film, the $z$ coordinate runs in direction of $\mathbf{n}$. The vectors $\mathbf{q}_i = q\mathbf{R}_{\pi/2}^i \cdot \mathbf{e}_x$, $(i = 0, 1)$, are two coplanar primitive reciprocal unit vectors of common modulus $q$, and $\mathbf{R}_{\pi/2}$ is an anticlockwise rotation matrix around the normal vector $\mathbf{n}$ by $\pi/2$. In previous work[30,34–36] we have fixed the modulus ($\nabla_\mathcal{A} q = 0$) and the pattern was a periodic square pattern of fixed period $a = 2\pi/q$.

Rendering the modulus, $q(\mathbf{r}_\mathcal{A})$, a function of action space ($\mathbf{r}_\mathcal{A} \in \mathcal{A}$) breaks the discrete translational symmetry. Here we

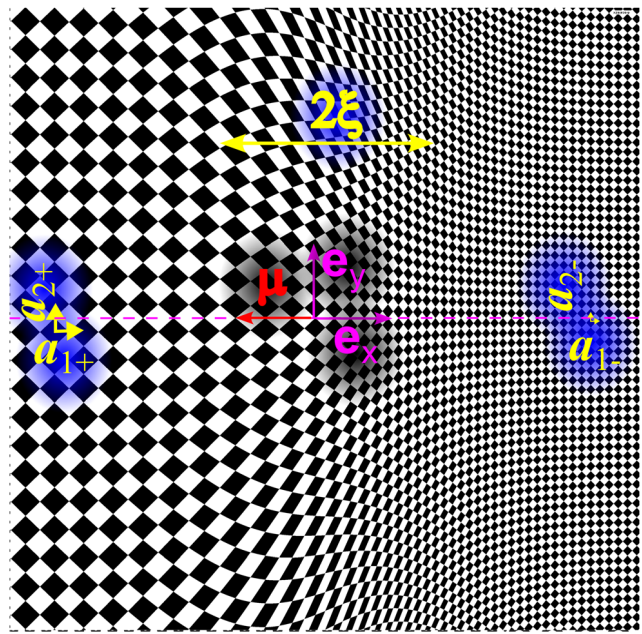

Fig. 2 Magnetic pattern (top view). The magnetization of the pattern follows eqn (1) with the modulus given by eqn (2). The larger lattice constant is $a_+ = 21$ μm, the magnification factor is $m = 3$ ($a_- = 7$ μm), and the transition length is $\xi = 75$ μm.

use a modulus

$$q(\mathbf{r}_\mathcal{A}) = \frac{2\pi/a_-}{1 + \dfrac{m-1}{2}(\tanh(\boldsymbol{\mu} \cdot \mathbf{r}_\mathcal{A}) + 1)}, \quad (2)$$

with a small and fixed morphing reciprocal vector $\boldsymbol{\mu} = -\mathbf{e}_x/\xi$. This continuously morphs the pattern within a transition region of width $2\xi$ from a periodic square pattern of period $a_- = a_+/m$ toward a periodic square pattern of a larger period $a_+$ (see Fig. 2).

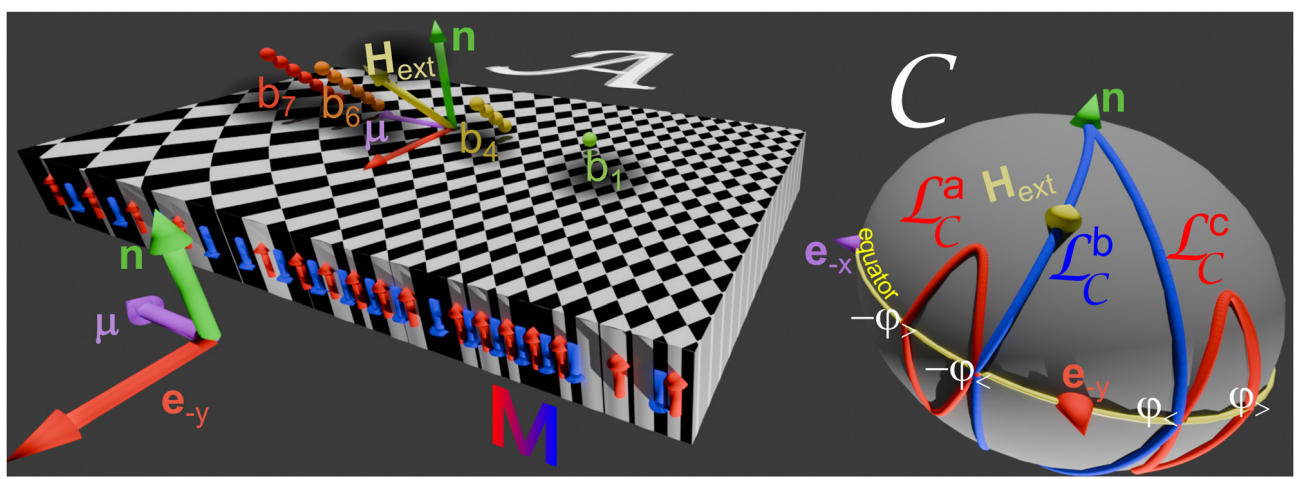

Fig. 1 Schematics of the setup. The pattern is a square pattern with alternating black and white domains that vary in the size of their lattice constant along the direction $\boldsymbol{\mu}$. The pattern is covered with a photoresist spacer layer of 1.3 μm thickness (not shown). Paramagnetic bipeds of length $b_n$ orient along the external magnetic field $\mathbf{H}_\mathrm{ext}$ and sediment with one foot on the spacer layer above the pattern. The external field adiabatically varies in control space $\mathcal{C}$ along a sorting loop $\mathscr{L}_\mathcal{C}^\mathrm{sort} = \mathscr{L}_\mathcal{C}^\mathrm{a} \times \mathscr{L}_\mathcal{C}^\mathrm{b} \times \mathscr{L}_\mathcal{C}^\mathrm{c}$ that crosses the equatorial plane at the characteristic loop angles $-\varphi_>$, $-\varphi_<$, $\varphi_<$, and $-\varphi_<$. The sorting loop is a concatenation of three loops, the red loops $\mathscr{L}_\mathcal{C}^\mathrm{a}$ and $\mathscr{L}_\mathcal{C}^\mathrm{c}$ winding in the mathematical negative sense and the blue loop $\mathscr{L}_\mathcal{C}^\mathrm{b}$ winding in the positive sense. The sorting loop lets the bipeds walk toward attractor lines perpendicular to the $\boldsymbol{\mu}$-direction that depends on the length of the biped.

In eqn (2) $m$ is the magnification factor between the asymptotic unit cell sizes. In the transition region, the distortion also breaks the periodicity of the pattern along the $\mathbf{e}_y$-direction (perpendicular to the morphing direction $\boldsymbol{\mu}$). In the $\mathbf{e}_y$-direction the unit cells are sheared with the amount of shearing increasing as one moves away form the central symmetry line located at $y = 0$.

A uniform time-dependent external field $\mathbf{H}_{\text{ext}}$ of constant magnitude $H_{\text{ext}} = 3.5$ kA m$^{-1}$ is superimposed to the non-uniform time-independent magnetic field generated by the pattern. The strong external field induces strong dipolar interactions between the colloidal particles which respond by self-assembling into rods of $n = 2$–$10$ particles with length $b_n = na$. The two ends of the rod are called the feet and when the rods start to walk we call them bipeds.

The orientation of the external field changes adiabatically along a closed loop $\mathscr{L}_{\mathscr{C}}^{\text{sort}}$ (Fig. 1). We call the set of possible external field orientations the control space $\mathscr{C}$, represented by a sphere in Fig. 1.

## 3. Periodic pattern

On top of a periodic pattern, where $\nabla_{\mathscr{A}} q = 0$, despite the field returning to its initial direction, single colloidal particles and bipeds can be topologically transported by one unit cell after completion of one loop.[30,33,35] The current section summarizes our previous work on the transport on periodic patterns[30,33,35] that already allowed us to separate bipeds by moving them into different directions but did not allow us to sort them to a final destination.

The orientation of the biped is locked to that of the external field with the northern foot being a magnetic north pole and the southern foot being a south pole. Let $\mathbf{b}_n$ denote the vector from the northern foot to the southern foot of a biped of length $b_n$. We call the three-dimensional vector space spanned by the end to end vectors $\mathbf{b}_n$ of the bipeds the transcription space $\mathscr{T}$. While the control space $\mathscr{C}$ has units of a magnetic field, transcription space $\mathscr{T}$ has units of a length. The loop $\mathscr{L}_{\mathscr{T}}^n = \frac{b_n}{H_{\text{ext}}} \mathscr{L}_{\mathscr{C}}$ is a transcription of the control loop $\mathscr{L}_{\mathscr{C}}$ onto a sphere of radius $b_n$ in transcription space $\mathscr{T}$. The transport occurs provided that the transcribed loop $\mathscr{L}_{\mathscr{T}}^n$ winds around specific fences $\mathscr{F}_{\mathscr{T}}^x$ and $\mathscr{F}_{\mathscr{T}}^y$ in transcription space.[30] These fences are parallel lines along the $\mathbf{e}_y$ and $\mathbf{e}_x$ directions equally spaced in $\mathscr{T}$ by the lattice period $a$ of the periodic pattern in action space $\mathscr{A}$. The central fence lines pass through the origin of transcription space $\mathscr{T}$. If the transcribed loop $\mathscr{L}_{\mathscr{T}}^n$ winds around the $\mathscr{F}_{\mathscr{T}}^x$ and $\mathscr{F}_{\mathscr{T}}^y$ lines with winding numbers $w_x^n(\mathscr{L}_{\mathscr{T}}^n) \in \mathbb{Z}$ and $w_y^n(\mathscr{L}_{\mathscr{T}}^n) \in \mathbb{Z}$ the biped is topologically transported by

$$\Delta \mathbf{r}_{\mathscr{A}} = a(w_x^n \mathbf{e}_x + w_y^n \mathbf{e}_y). \tag{3}$$

Fig. 3 shows two transcriptions $\mathscr{L}_{\mathscr{T}}^3$ and $\mathscr{L}_{\mathscr{T}}^6$ of the same loop $\mathscr{L}_{\mathscr{C}}$ with different winding numbers around the fences $\mathscr{F}_{\mathscr{T}}^x$ and $\mathscr{F}_{\mathscr{T}}^y$ in transcription space $\mathscr{T}$. The winding number $w_x^3 = 0$ and hence a biped of length $b_3$ will not be transported by the loop $\mathscr{L}_{\mathscr{C}}$, while the winding number $w_x^6 = 1$ and therefore a biped of length $b_6$ will be transported by the displacement $\Delta \mathbf{r}_{\mathscr{A}} = a\mathbf{e}_x$.

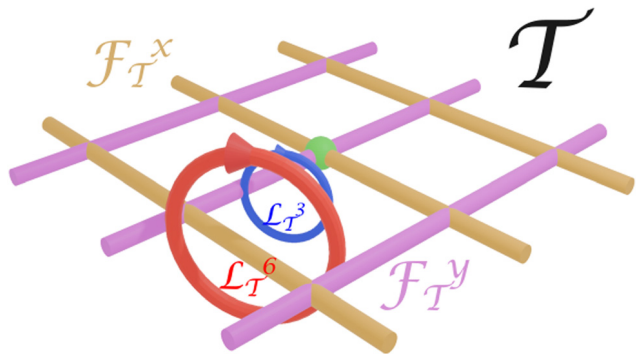

Fig. 3 Transcriptions of the same loop $\mathscr{L}_{\mathscr{C}}$ in control space $\mathscr{C}$ to the loops $\mathscr{L}_{\mathscr{T}}^3$ and $\mathscr{L}_{\mathscr{T}}^6$ in transcription space $\mathscr{T}$ corresponding to the biped sizes $b_3$ and $b_6$ with different winding numbers $w_x^n$ around the fences $\mathscr{F}_{\mathscr{T}}^x$. The green sphere marks the origin of transcription space $\mathscr{T}$ and the loops $\mathscr{L}_{\mathscr{T}}^3$ and $\mathscr{L}_{\mathscr{T}}^6$ lie on corresponding spheres centered at the origin.

## 4. Non-periodic pattern

Due to the topological nature of the transport, eqn (3) remains valid for small deformations of the periodic pattern. We have shown that eqn (3) is invariant under rotations[39] and under conformal transformations[40] of the pattern and invariant under modulations of the symmetry phase of patterns with three fold symmetry.[32] We therefore expect the displacement on top of a deformed non-periodic four fold pattern with varying lattice constant $\nabla_{\mathscr{A}} q \neq 0$ to be expressed by the same eqn (3), provided that the deformation is locally small. That is:

$$\Delta \mathbf{r}_{\mathscr{A}} = a(\mathbf{r}_{\mathscr{A}})(w_x^n(\mathbf{r}_{\mathscr{A}})\mathbf{e}_x + w_y^n(\mathbf{r}_{\mathscr{A}})\mathbf{e}_y), \tag{4}$$

with $a(\mathbf{r}_{\mathscr{A}})$ the local value of the lattice constant at the position $\mathbf{r}_{\mathscr{A}}$, and with local fences lines in transcription space $\mathscr{F}_{\mathscr{T}}^x(\mathbf{r}_{\mathscr{A}})$ and $\mathscr{F}_{\mathscr{T}}^y(\mathbf{r}_{\mathscr{A}})$ now being parallel lines along the $\mathbf{e}_y$ and $\mathbf{e}_x$ direction equally spaced by the local lattice constant $a(\mathbf{r}_{\mathscr{A}})$ of the non-periodic pattern in action space $\mathscr{A}$. The loop $\mathscr{L}_{\mathscr{T}}^n$ is independent of the location $\mathbf{r}_{\mathscr{A}}$ of the biped on the pattern. However the fences change with the biped location $\mathbf{r}_{\mathscr{A}}$ and so do the winding numbers of the fixed loop around the varying fences. It is for this reason that the winding numbers and thus the displacement of a fixed loop $\mathscr{L}_{\mathscr{C}}$ causes different transport for different biped length $b_n$ as well as for different locations $\mathbf{r}_{\mathscr{A}}$.

## 5. Sorting loop

We test the validity of eqn (4) using a sorting loop $\mathscr{L}_{\mathscr{C}}^{\text{sort}} = \mathscr{L}_{\mathscr{C}}^a \times \mathscr{L}_{\mathscr{C}}^b \times \mathscr{L}_{\mathscr{C}}^c$ being the concatenation of three fundamental loops $\mathscr{L}_{\mathscr{C}}^a$, $\mathscr{L}_{\mathscr{C}}^b$, and $\mathscr{L}_{\mathscr{C}}^c$. We present $\mathscr{L}_{\mathscr{C}}^{\text{sort}}$ in Fig. 1. We call a fundamental loop a loop that crosses the equator of control space twice: once at time $t_\uparrow$ up from the southern hemisphere to the northern hemisphere at an external field $\mathbf{H}_{\text{ext}}(t_\uparrow)$ and once at time $t_\downarrow$ down from the northern hemisphere to the southern hemisphere at an external field $\mathbf{H}_{\text{ext}}(t_\downarrow)$. The angle of the crossing field direction to the reciprocal unit vector $\mathbf{q}_1$ is given by $qH_{\text{ext}} \sin(\psi_{\uparrow\downarrow}) = (\mathbf{q}_1 \times \mathbf{H}_{\text{ext}}(t_{\uparrow\downarrow})) \cdot \mathbf{n}$. We choose a loop $\mathscr{L}_{\mathscr{C}}^{\text{sort}}$ with the three pairs of equator crossing angles $\psi_\uparrow^a, \psi_\downarrow^a, \psi_\uparrow^b, \psi_\downarrow^b, \psi_\uparrow^c$, and $\psi_\downarrow^c$ that may be described by the two angles $\varphi_<$ and $\varphi_>$ as indicated in Table 1 (see also Fig. 1). Due to the symmetric choice of the angles, the winding number around the $\mathcal{F}_\mathcal{T}^y$-fences vanishes irrespective of the length of the biped $w_y^n(\mathcal{L}_\mathcal{C}^{sort}) = 0$ and we expect no topological transport along the $y$-direction.

Table 1 Properties of the sorting loop

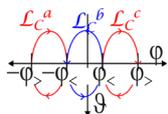

|  $\mathcal{L}_\mathcal{C}^a$ | |  $\mathcal{L}_\mathcal{C}^b$ | |  $\mathcal{L}_\mathcal{C}^c$ | |
| --- | --- | --- | --- | --- | --- |
| $\psi_\uparrow^a$ | $\psi_\downarrow^a$ | $\psi_\downarrow^b$ | $\psi_\uparrow^b$ | $\psi_\uparrow^c$ | $\psi_\downarrow^c$ |
| $-\varphi_>$ | $-\varphi_<$ | $-\varphi_<$ | $\varphi_<$ | $\varphi_<$ | $\varphi_>$ |

## 6. Experimental results

We define a parameter $\lambda$ to characterize the sorting loop as $\lambda(\mathcal{L}_\mathcal{C}^{sort}) = \sin\varphi_>/(2\sin\varphi_<)$. Due to the symmetric choice of the sorting loop, $\lambda$ is the only relevant loop parameter determining the winding numbers. We have performed experiments using sorting loops with $\lambda = 0.67$, $0.76$, $0.84$, and $1.17$ that sort bipeds of size $b_3$–$b_8$ with single particle diameter $d = 2.8$ μm on top of the pattern depicted in Fig. 2. Fig. 4 shows an overlay of microscopy images of the pattern covered with bipeds subject to a sorting loop with $\lambda = 0.84$ at different times together with trajectories of those bipeds obtained via particle tracking. Single particles, $n = 1$ (red), and doublets $n = 2$, are traveling one unit cell per loop in the positive $x$-direction, i.e., toward the unit cells of smaller size $a_-$. The travel direction of larger bipeds ($n > 2$) depends on the location on the pattern. For each biped size $b_n$ there is a set of attractor lines $r_\mathcal{A}^{*n}$ along the $y$-direction separated by separatrixes $r_\mathcal{A}^{s,n}$. For initial positions $r_\mathcal{A}$ in the region between two adjacent separatrixes all bipeds of size $b_n$ are attracted toward the attractor line $r_\mathcal{A}^{*n}$ between both seperatrixes. On either side of the attractor line after each application of the sorting loop the bipeds perform a step by one, two or more unit cells closer to the attractor line. At the attractor line after each application of the sorting loop, bipeds step from one side of the attractor line toward the other side of the line. This essentially makes the bipeds stop to move after reaching the attractor line.

Far away from the symmetry line, $y = 0$, the unit cells of the pattern are sheared. A shear is a superposition of an extension and a rotation. The rotation of cells in action space also rotates the direction of the fence lines in transcription space. This also causes transport into the $y$-direction, of bipeds residing in the transition region further away from the symmetry line (see e.g. the $b_5$- and the $b_6$-trajectories in the lower part of Fig. 4).

Because the attractor line depends on the length $b_n$ of the biped, all bipeds are essentially sorted to their individual attractor line. In Fig. 4 we also see a non-sorted $b_6$-biped to the right of the $b_6$-separatrix walking away from the separatrix toward a second $b_6$-attractor line, with $a(r_\mathcal{A}^{*n}) < a_-$, that does not exist on the current pattern since the lattice constant does not shrink below the minimum lattice constant of $a_-$.

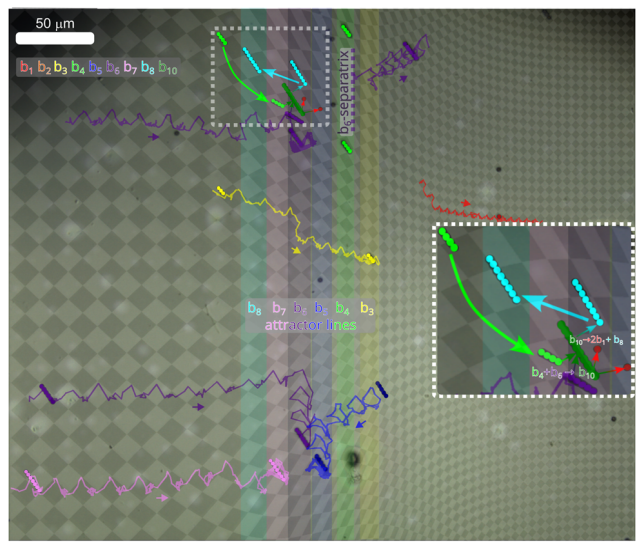

Fig. 4 Sorting experiments. Overlay of several microscopy images of bipeds at different times above the sorting pattern, together with trajectories of bipeds of different length $b_n$ subject to the repeated sorting loop with sorting parameter $\lambda = 0.84$. The trajectories are colored according to the biped length as indicated in the image. The magnified dotted square region shows a biped reaction $b_4 + b_6 \to b_{10} \to b_8 + 2b_1$ and the resorting of the resulting $b_8$-biped. Supplementary movie 1 (ESI†) provides a dynamic impression of the sorting process. Supplementary movie 2 (ESI†) shows the sorting of a $b_7$-biped that is missing in the experiment shown in this figure.

Eventually a walking biped of length $b_m$, can collide with a sorted and therefore no longer walking biped of larger length $b_n$ with $n > m$ to form a longer biped of length $b_{n+m}$. The attractor line for the $b_{n+m}$-biped is different from both attractor lengths of the bipeds prior to the collision. Hence, they are sorted to their new attractor line after the collision. The inset of Fig. 4 is a magnification of the dotted squared region of the main figure. There we see a $b_6$-biped already sorted to its attractor line. A not yet sorted $b_4$-biped while walking to its attractor line collides with the sorted $b_6$ biped an initiates an addition reaction $b_4 + b_6 \to b_{10}$. The resulting $b_{10}$-biped loses two single colloidal particles, i.e. $b_{10} \to b_8 + 2b_1$ leaving a non-sorted $b_8$ behind that is then sorted toward its attractor line by moving into the $-x$ direction. Supplementary movie 1 (ESI†) shows the sorting of bipeds of length $b_3$, $b_4$, $b_5$, $b_6$ and $b_8$ and the resorting after a biped addition reaction of Fig. 4. Supplementary movie 2 (ESI†) shows the sorting of bipeds of length $b_7$ toward its attractor line. Note that Supplementary movie 1 and 2 (ESI†) are speeded up by a factor of 50 and 70, respectively.

## 7. Winding number

The winding number around the family of $\mathcal{F}_\mathcal{T}^x$ fence lines depends on the sorting loop parameter $\lambda$ and the biped length measured in units of the local lattice constant $\beta(r_\mathcal{A}) = b_n \sin\varphi_</a(r_\mathcal{A})$. The winding number can be computed using the construction of Fig. 3 to our sorting loop $\mathcal{L}^{sort}$ and it reads:

$$w_x^n(\lambda, \beta) = 1 + 4[\beta(\mathbf{r}_\mathcal{A})] - 2[2\lambda\beta(\mathbf{r}_\mathcal{A})] \quad (5)$$

where the square brackets $[x]$ denote the floor- (ceiling-) function for $x \gtrless 0$. The winding number $w_x^n(\lambda, \beta)$ around the $\mathcal{F}_{\mathcal{F}}^x$ fence lines is an odd integer that can be smaller or larger than zero. In Fig. 5 we plot the signum of the winding number as a function of $\beta(\mathbf{r}_{\mathcal{A}})$ and $\lambda$. The border between positive ($w_x^n = 1$) and negative ($w_x^n = -1$) winding number is a line given by

$$\lambda^c = \frac{2[\beta] + 1}{2\beta} \quad (6)$$

that oscillates around the (dashed white) line $\lambda = 1$. The walking direction is toward the positive $x$-direction in the bright gray regime and toward the negative $x$-direction in the dark gray regime. Small bipeds always walk toward the right. If the gradient of the lattice constant points in the negative $x$-direction $\nabla_{\mathcal{A}} a(\mathbf{r}_{\mathcal{A}}) \cdot \mathbf{e}_x < 0$, as in our pattern, the parameter $\beta(\mathbf{r}_{\mathcal{A}})$ will increase with each walking step to the right. The walking direction changes at the position $\beta(\mathbf{r}_{\mathcal{A}}^{*n})$ when the bipeds cross from the bright gray ($w_x < 0$) to the line between the dark ($w_x > 0$) and the bright gray region at a negative slope of the line $\lambda^c(\beta)$. Negative slopes of the line $\lambda^c(\beta)$ between positive and negative winding numbers are attractors of bipeds that are classified by the integer $[\beta]$. Sections with positive slopes are separatrices between regions where bipeds are attracted toward different sections with negative slope. For $\lambda = 1$ there are an infinite number of attractor sections while for $\lambda \neq 1$ the number of attractor sections is finite.

We check our sorting experiments by measuring the biped length $b_n$, the attractor positions $\mathbf{r}_{\mathcal{A}}^{*n}$ and the local unit cell size $a(\mathbf{r}_{\mathcal{A}}^{*n})$ of the attractor line for all different biped sizes and loop parameters $\lambda$. From these measurements we compute $\beta(\mathbf{r}_{\mathcal{A}}^{*n}) = b_n \sin \varphi_< / a(\mathbf{r}_{\mathcal{A}}^{*n})$. Experimental pairs of $\beta(\mathbf{r}_{\mathcal{A}}^{*n})$ and $\lambda$ are then plotted as star symbols into Fig. 5. For all different lengths of bipeds $b_n$, $n = 3,\ldots,8$ the experimental attractor points fit well on the theoretical prediction of eqn (6). Bipeds attracted to the second attractor line with $[\beta] = 1$ start walking from initial locations $\mathbf{r}_{\mathcal{A}}^{i,n}$, $i = 1,\ldots,k$ in the direction opposite to the separatrix location $\mathbf{r}_{\mathcal{A}}^{s,n}$. In Fig. 5 we plot the corresponding pairs $\lambda$, and $\min_{i=1}^{k} \beta(\mathbf{r}_{\mathcal{A}}^{i,n})$, that lie closest to the separatrix for each value of $\lambda$ (solid circle). These points are upper bounds for the separatrix location. All initial locations of bipeds approaching the $[\beta] = 1$ attractor line (open circles) lie in the bright gray region of negative winding number $w_x^n < 0$ consistent with the theoretical prediction. Their very different separations from the theoretical separatrix is due to the random initial placements of the set of bipeds on the pattern.

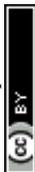

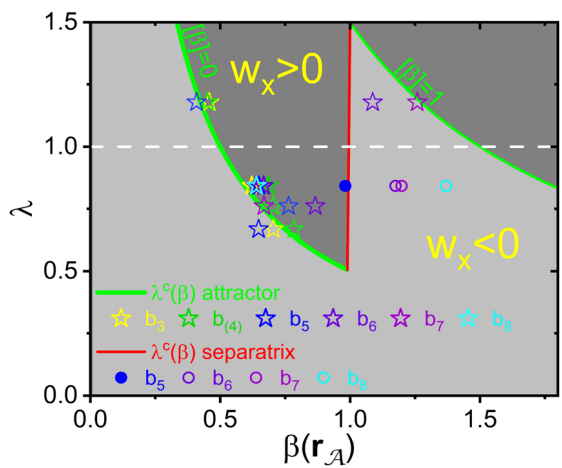

Fig. 5 The winding number $w_x^n$, eqn (5), as a function of $\beta(\mathbf{r}_{\mathcal{A}}) = b_n \sin \varphi_< / a(\mathbf{r}_{\mathcal{A}})$ and $\lambda$. If the gradient of the lattice constant points in the negative $x$-direction $\nabla_{\mathcal{A}} a(\mathbf{r}_{\mathcal{A}}) \cdot \mathbf{e}_x < 0$ negative slopes of the line, eqn (6), between positive and negative winding numbers are attractors for bipeds, while sections with positive slopes are separatrices between regions where bipeds are attracted toward different sections with negative slope. The star symbols correspond to final experimental positions of the bipeds, the circles are initial experimental positions beyond the first separatrix with bipeds walking toward the $[\beta] = 1$ attractor line. The solid circle is most left initial position observed among all those initial positions.

## 8. Discussion and conclusion

Our experimental data is in good agreement with the predictions of eqn (6). Therefore by choosing a specific sorting loop, bipeds walk to their attractor position without any need to enforce their proper walk using a feedback mechanism. The decision to walk to the proper place under the periodic external driving is made internally by the bipeds. This is particularly prominent when an addition reaction $b_n + b_m \to b_{n+m}$ changes the length of two colliding bipeds toward a third length and the longer length biped is automatically resorted. For an attractor line family with the same integer classification $[\beta]$, the attractor line position evolves continuously with the length $b_n$ of the bipeds. Using attractor lines with family classification $[\beta] > 0$ it is possible to separate larger length bipeds from shorter biped attractor destinations even further, when choosing a driving loop with parameter $\lambda \neq 1$ such that the larger biped is part of the family with classification $[\beta] > 0$, while the shorter is not.

We have sorted bipeds of length $b_3$–$b_8$. One may ask what happens to bipeds of shorter and longer length. Single colloidal particles and $b_2$ bipeds are both shorter than the smallest unit cell on the pattern. Therefore the transcription of control space into transcription space for these two biped lengths are spheres of radius 2.8 μm and 5.6 μm. Both transcription spheres are smaller than the smallest unit cell (the spacing of the fences in Fig. 3). The winding number of the transcribed side loops $w_x(\mathscr{L}_{\mathcal{F}}^{na}) = 0 = w_x(\mathscr{L}_{\mathcal{F}}^{nc})$ vanish anywhere on the pattern, and the winding number of the transcribed central loops $w_x(\mathscr{L}_{\mathcal{F}}^{nb}) = 1$ is constant and positive anywhere on the pattern. Single colloidal particles and $b_2$ bipeds therefore walk toward the right searching for putative attractor lines smaller than the smallest unit cell size $a_-$, which they can never find on this pattern. Single colloidal particles and $b_2$ bipeds are thus never sorted.

In the experiments there are no bipeds of size $b_{10}$ or larger before starting the loops. Such long bipeds only occur due to irreversible additions of smaller bipeds to the longer biped. For an addition reaction to occur, smaller bipeds must collide with each other. This happens predominantly in the attractor regions of the $b_3$–$b_8$ bipeds because it is in this region that smaller bipeds of different length counter propagate. Long bipeds are therefore found mostly on the right half of the pattern, which is a location already to the right of the first separatrices of these bipeds. Like the single colloidal particles and $b_2$ bipeds, they start to search for an attractor line smaller than 7 μm that does not exist on the pattern. Ultimately they are not sorted for the very same reason as the single colloidal particles and $b_2$ bipeds.

The probability of collisions in the attractor regions of the $b_3$–$b_8$ bipeds is also the reason one cannot increase the concentration of colloids. Colloids will collide in this region to form large bipeds that leave the pattern and only a small amount of sorted bipeds similar to what we see in our experiments presented remain sorted on the pattern.

We define the efficiency of sorting as the probability that a biped $b_n$ is transported to the correct attractor line $\lambda^c(\beta)$ without colliding with another biped to form a biped of a different length prior to the sorting. At our low concentration of particles the efficiency of the sorting is roughly 90%. Note that sorting is not the same as separation. Attractor lines for bipeds of different lengths $b_n$ and $b_{n+dn}$ are separated by a distance that is the larger the smaller is the morphing reciprocal vector $\mu$. If we define the efficiency of separation as the differential of separation of attractor lines with respect to the variation biped length $e_{\text{separate}} = \frac{dr_{\mathcal{A}}^{*n}}{dn} \propto \frac{\xi}{m-1}$, then the efficiency of separation is proportional to the correlation length $\xi$ of the pattern. A third efficiency is the time efficiency, i.e. the time we need to sort the bipeds. The deeper the colloidal potential, the faster we can modulate and still be in the adiabatic regime. The colloidal potential is deeper the farther we are separated with the loop in transcription space from the fence lines. This is best done amongst the family of topologically equivalent loops by choosing loops that avoid spending significant arclength of their path near the equator. In our experiments we use a period of $T = 20$ s per fundamental loop. The pattern and the loop are optimized with respect to the task of sorting bipeds to different predefined locations given the constraints of also being able to simultaneously observe all bipeds.

In conclusion we placed paramagnetic colloidal particles on top of a square-like magnetization pattern with alternating domains of magnetization with a locally varying unit cell size and apply a sorting loop $\mathscr{L}_{\mathscr{C}}^{\text{sort}}$ that lets bipeds assembled via dipolar attraction automatically sort themselves to attractor lines that depend on the length of the bipeds. In contrast to separation techniques like electrophoresis, the biped length do not separate as a function of time but all bipeds walk simultaneously to their attractor final destination that varies with the length of the bipeds.

## Author contributions

ARP, DdlH, & TMF designed and performed the experiment and wrote the manuscript with input from all the other authors. PK, MU & FS produced the magnetic film. SA, & ArE performed the fabrication of the micromagnetic patterns within the magnetic thin film.

## Data availability

The data supporting this article have been included as part of the ESI.†

## Conflicts of interest

There are no conflicts to declare.

## Acknowledgements

This work is funded by the Deutsche Forschungsgemeinschaft (DFG, German Research Foundation) under project number 531559581. DdlH acknowledges support via the Heisenberg program of the DFG via project number HE 7360/7-1. P. K., F. S., and M. U. acknowledge financial support from the National Science Centre Poland through the OPUS funding (Grant No. 2019/33/B/ST5/02013). S. A. acknowledges funding by a PhD scholarship of Kassel University. A. E., S. A. acknowledges support by the Deutsche Forschungsgemeinschaft (DFG, German Research Foundation) under project numbers 514858524, INST 159/139-1 FUGG, INST 159/94-1 FUGG, and INST 159/95-1 FUGG.

## Notes and references

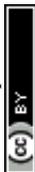